\documentstyle[aps,epsf,multicol,graphicx]{revtex}
\begin{document}
\title{Impurity in a Maxwellian Unforced Granular Fluid}
\author{E.~Ben-Naim$^1$ and P.~L.~Krapivsky$^2$}
\address{$^1$Theoretical Division and Center for Nonlinear Studies,
Los Alamos National Laboratory, Los Alamos, NM 87545, USA}
\address{$^2$Center for Polymer Studies and Department of Physics,
Boston University, Boston, MA 02215, USA}
\maketitle
\begin{abstract}
  We investigate velocity statistics of an impurity immersed in a
  uniform granular fluid. We consider the cooling phase, and obtain
  scaling solutions of the inelastic Maxwell model analytically.
  First, we analyze identical fluid-fluid and fluid-impurity collision
  rates.  We show that light impurities have similar velocity
  statistics as the fluid background, although their temperature is
  generally different. Asymptotically, the temperature ratio increases
  with the impurity mass, and it diverges at some critical mass.
  Impurities heavier than this critical mass essentially scatter of a
  static fluid background.  We then analyze an improved inelastic
  Maxwell model with collision rates that are proportional to the
  {\it average} fluid-fluid and fluid-impurity relative velocities.
  Here, the temperature ratio remains finite, and the system is always
  in the light impurity phase.  Nevertheless, ratios of sufficiently
  high order moments $\langle v^n_{\rm impurity}\rangle/\langle
  v^n_{\rm fluid}\rangle$ may diverge, a consequence of the
  multiscaling asymptotic behavior.  
\end{abstract}
\begin{multicols}{2}

\section{Introduction}
Granular media are typically polydisperse. For example, sand and
grains have a broad range of particle sizes and shapes. Such granular
mixtures exhibit size segregation, a ubiquitous collective phenomena
that underlies diverse processes including for example production and
transport of powders in industry, sand dunes propagation and volcanic
flows in geophysics \cite{jcw,csc}. The ``Brazil Nut'' problem where
an impurity is immersed in a uniform granular media is an extreme
realization of a granular mixture as it corresponds to the vanishing
volume fraction limit of a binary mixture. While this problem has been
extensively studied, dynamics of such impurities are not fully
understood \cite{rsps,al,jmp,drc,kjn,dj,hql,th}.

Understanding the velocity statistics of granular mixtures is a
necessary step in describing multiphase granular flows.  Experimental
and theoretical studies show that in general, different components of
a granular mixture are characterized by different typical speeds,
i.e., granular temperatures are usually distinct \cite{wp}. However,
analytical treatment of this case is difficult given the coupling
between the different components of the mixture.  An additional
complication arises from the large number of parameters, including
several restitution coefficients, governing the dynamics.

In contrast, the impurity problem is more amenable to analytical
treatment.  First, the impurity is directly enslaved to the fluid and
second, there are fewer parameters \cite{sd}.  In this paper, we study
the impurity problem in the framework of the Maxwell
model\cite{max,true} that assumes that the collision rate is uniform,
i.e., independent of the relative velocity of the colliding
particles. This simplifies the Boltzmann collision operator and thence
this model is widely used in kinetic theory\cite{true,e,bob}.  The
Maxwell model is analytically tractable even when the collisions are
inelastic as shown in a number of recent studies of uniform and
polydisperse granular gases \cite{bk,bcc,bc,bmp,kb,eb,mp,santos}.

We obtain analytic results for the velocity distributions valid for
arbitrary spatial dimension and collision parameters. We consider two
versions of the Maxwell model.  In the first, termed the Inelastic
Maxwell Model (IMM), the collision rates are completely independent of
the relative velocities of the colliding particles.  In the second,
termed the Improved Inelastic Maxwell Model (IIMM), the collision
rates are proportional to the {\it average} relative velocity of the
colliding particles.

In the IMM, there are two phases separated by a critical impurity
mass $m_*(r_p,r_q)$, determined by the restitution coefficients $r_p$ and
$r_q$ characterizing fluid-fluid and fluid-impurity particle collisions,
respectively.  When the impurity mass $m$ is smaller than the critical mass,
$m<m_*$, different temperatures characterize the fluid and the impurity. The
impurity velocity distribution exhibits similar characteristics as does the
fluid; in particular, the same exponent governs the algebraic decay of the
large velocity tail.  Asymptotically, the impurity to fluid temperature ratio
diverges as the critical mass is approached and is infinite when $m\geq m_*$.
In this regime, the impurity essentially scatters of an ensemble of static
fluid particles. The governing Lorentz-Boltzmann equation simplifies
considerably.  In general, moments of the velocity distribution exhibit
multiscaling asymptotic behavior and the velocity distribution consists of
replicas of the initial conditions. On the other hand, this phase transition
is suppressed in the IIMM. The temperature ratio remains finite and the
system is always in the light impurity phase.  Nevertheless, secondary
transitions affecting large velocity moments remain. 

The rest of this paper is organized as follows. In Sec.~II, we
describe the IMM and present the Boltzmann equation for the fluid
velocity distribution and the Lorentz-Boltzmann equation for the
impurity velocity distribution.  We then determine the temperature and
show that a phase transition occurs.  The light impurity phase is
analyzed first and then the heavy impurity phase, emphasizing the
behavior of the velocity distribution and its moments.  In Sec.~III,
we discuss the IIMM and show that the system always remains in the
light impurity phase.  We conclude with a summary.

\section{The Inelastic Maxwell Model}

We study dynamics of a single impurity particle in a uniform
background of identical inelastic spheres. Without loss of generality,
we set the mass of the fluid particles to unity, while the mass of the
impurity is denoted by $m$.  Particles interact via binary collisions
that lead to exchange of momentum along the impact direction.
Collisions between two fluid particles are characterized by the
collision parameter $p$, and collisions between the impurity and any
fluid particle are characterized by the collision parameter $q$.  When
a particle of velocity ${\bf u}_1$ collides with a fluid particle of
velocity ${\bf u}_2$ its post-collision velocity ${\bf v}_1$ is given
by
\begin{eqnarray}
\label{rulep}
{\bf v}_{1}&=&{\bf u}_{1}-(1-p)\,({\bf g}\cdot {\bf n})\,{\bf n},\\
\label{ruleq}
{\bf v}_{1}&=&{\bf u}_{1}-(1-q)\,({\bf g}\cdot {\bf n})\,{\bf n},
\end{eqnarray}
with ${\bf g}={\bf u}_1-{\bf u}_2$ the relative velocity and ${\bf n}$
the unit vector parallel to the impact direction. The first equation
gives the velocity of a fluid particle and the second that of an
impurity.  The collision rules (\ref{rulep})--(\ref{ruleq}) are
derived by employing momentum conservation combined with the fact that
in an inelastic collision, the component of the relative velocity
parallel to the impact direction is reduced by a factor equal to the
fluid-fluid (impurity-fluid) restitution coefficient $r_p$ ($r_q$).
The restitution coefficients are related to the collision parameters
via
\begin{eqnarray}
\label{rp}
r_p&=&1-2p,\\
\label{rq}
r_q&=&m-(m+1)\,q.
\end{eqnarray}
Since the restitution coefficients obey $0\leq r\leq 1$, the collision
parameters satisfy $0\leq p\leq 1/2$ and \hbox{${m-1\over m+1}\leq
q\leq {m\over m+1}$}.  The energy dissipated in each collision is
equal to $A({\bf g}\cdot {\bf n})^2$, with $A=2p(1-p)$ and
$A=m(1-q)[2-(m+1)(1-q)]$ for fluid-fluid and impurity-fluid
collisions, respectively.

We consider a collision process where random pairs of particles
undergo inelastic collisions with a random impact direction.  This
inelastic Maxwell model (IMM) is described by a Boltzmann equation
with a uniform collision rate. The model is a straightforward
generalization of the classical Maxwell model.  Specifically, Maxwell
showed \cite{max} that for elastic ``Maxwell molecules'' interacting
via a repulsive $r^{-2(d-1)}$ potential in $d$ spatial dimensions, the
collision rate does not depend on the magnitude of the relative
velocity.

Let $P({\bf v},t)$ and $Q({\bf v},t)$ be the normalized velocity
distributions of the background and the impurity, respectively. In
this collision process, no correlations develop and the governing
Boltzmann equation and the Lorentz-Boltzmann equation are
\begin{eqnarray}
\label{bep}
{\partial P({\bf v},t)\over \partial t}
&=&\int d{\bf n}\int d{\bf u}_1\,P({\bf u}_1,t)
\int d{\bf u}_2\,P({\bf u}_2,t) \\
&\times& \left\{\delta\Big[{\bf v}-{\bf u}_1+(1-p)({\bf g}
\cdot {\bf n}){\bf n}\Big]-\delta({\bf v}-{\bf u}_1)\right\},\nonumber
\\
\label{beq}
{\partial Q({\bf v},t)\over \partial t}
&=&\int d{\bf n}\int d{\bf u}_1\,Q({\bf u}_1,t)
\int d{\bf u}_2\,P({\bf u}_2,t) \\
&\times&
\left\{\delta\Big[{\bf v}-{\bf u}_1+(1-q)({\bf g}\cdot {\bf n}){\bf n}\Big]
-\delta({\bf v}-{\bf u}_1)\right\}.\nonumber
\end{eqnarray}
The angular integration over the impact direction is normalized to
unity, $\int d{\bf n}=1$.  The collision integrals directly reflect
the collision rules.  

In writing Eqs.~(\ref{bep})--(\ref{beq}), fluid-fluid and
impurity-fluid collision rates were assumed to be the same and were
set to unity for convenience.  Hence the average number of collisions
experienced by a particle equals time.  The results detailed in this
section are exact for a mean-field collision process where,
irrespective of their type, random pairs of particles are chosen to
undergo inelastic collisions according to
(\ref{rulep})--(\ref{ruleq}). A more realistic collision rates model
is treated in Sec.~III.

The Boltzmann and Lorentz-Boltzmann equations simplify in Fourier
space.  Thence we use the Fourier transforms of the velocity
distribution functions,
\begin{eqnarray}
\label{fkt-def}
F({\bf k},t)&=&\int d{\bf v}\,e^{i{\bf k}\cdot {\bf v}}\,P({\bf v},t),\\
\label{gkt-def}
G({\bf k},t)&=&\int d{\bf v}\,e^{i{\bf k}\cdot {\bf v}}\,Q({\bf v},t).
\end{eqnarray}
The convolution structure of Eqs.~(\ref{bep})--(\ref{beq}) implies that the
collision terms factorize
\begin{eqnarray}
\label{fkt-rate}
{\partial \over \partial t}F({\bf k},t)+F({\bf k},t)
&=& \int d{\bf n}\,F\left[{\bf k}-{\bf p},t\right]
F\left[{\bf p},t\right],\\
\label{gkt-rate}
{\partial \over \partial t}G({\bf k},t)+G({\bf k},t)
&=& \int d{\bf n}\,G\left[{\bf k}-{\bf q},t\right]
F\left[{\bf q},t\right],
\end{eqnarray}
with ${\bf p}=(1-p)\,({\bf k}\cdot{\bf n})\,{\bf n}$ and ${\bf
q}=(1-q)\,({\bf k}\cdot{\bf n})\,{\bf n}$. These equations reflect the
momentum transfer occurring during collisions. They also hint that
the model is analytically tractable.  For example, expansion in
powers of the wave number shows that moments of the velocity
distributions obey closed hierarchies of evolution equations.

\subsection{The Temperature}

We study the freely evolving case where in the absence of energy input the
system ``cools'' indefinitely \cite{pkh,gz}.  The fluid temperature is
defined as the average square of the velocity: $T={1\over d}\int d{\bf v}\,
v^2 P({\bf v},t)$ with $v\equiv |{\bf v}|$.  {}From the Boltzmann equation
(\ref{bep}), the fluid temperature evolves according to ${d\over
  dt}T=-\lambda T$ with \hbox{$\lambda=2p(1-p)\int d{\bf n}\, n_1^2$}.  The
angular integration is readily performed, $\int d{\bf n}\, n_1^2=1/d$, by
using symmetry and the identity $n_1^2+\ldots+n_d^2=1$.  Hence the fluid
temperature satisfies
\begin{equation}
\label{tl}
{d \over dt}T=-{2p(1-p)\over d}\,T,
\end{equation}
implying that the fluid temperature decays exponentially with time
\begin{equation}
T(t)=T_0\,e^{-2p(1-p)t/d}.
\end{equation}

Similarly, from the Lorentz-Boltzmann equation (\ref{beq}) one can
obtain the governing equation for the impurity temperature
$\Theta(t)={1\over d}\int d{\bf v}\, v^2 Q({\bf v},t)$.  One finds
that the impurity temperature is coupled to the fluid temperature via
a simple linear rate equation
\begin{equation}
\label{theta-eq}
{d \over dt}\Theta=-{1-q^2\over d}\,\Theta +{(1-q)^2\over d}\, T.
\end{equation}
The solution to this equation is a linear combination of two exponentials
\begin{equation}
\label{theta-sol}
\Theta(t)=\left(\Theta_0-c\,T_0\right)\,e^{-(1-q^2)t/d}
+c\,T_0\,e^{-2p(1-p)t/d},
\end{equation}
with the constant
\begin{equation}
\label{c}
c={(1-q)^2\over 1-q^2-2p(1-p)}.
\end{equation}

Therefore, there are two different regimes of behavior. When $1-q^2>2p(1-p)$,
the impurity temperature is proportional to the fluid temperature
asymptotically, ${\Theta(t)\over T(t)}\to c$ as $t\to \infty$.  In the
complementary region $2p(1-p)>1-q^2$ an extreme violation of equipartition
occurs, as the ratio of the fluid temperature to the impurity temperature
vanishes. The impurity is very energetic compared with the fluid and it
practically sees a static fluid.  {}From Eq.~(\ref{c}) we find that at the
transition point $q=\sqrt{1-2p(1-p)}$.  Employing relations
(\ref{rp})--(\ref{rq}) between the restitution coefficients and the collision
parameters we obtain the critical mass
\begin{equation}
\label{mstar}
m_*={r_q+\sqrt{(1+r_p^2)/2}\over 1-\sqrt{(1+r_p^2)/2}}.
\end{equation}
The heavy impurity phase arises when the impurity is a bit heavier than the
fluid particles: Even when the fluid-fluid collisions are completely
inelastic ($r_p=0$), the critical mass satisfies $m_*>1+\sqrt{2}$.  For
weakly dissipative fluids ($r \to 1$), the critical mass diverges,
$m_*\propto (1-r_p)^{-1}$ (see Fig.~1).

Note now a few features of the light impurity phase.  First, Eq.~(\ref{c})
generalizes the elastic fluid ($p=0$) result $c=(1-q)/(1+q)$\cite{p,mp1}.
That result was actually established for a hard sphere fluid, so at least
asymptotically both the IMM and the improved model predict the same
impurity temperature when the fluid is elastic.  Further, the initial
impurity temperature becomes irrelevant and the impurity is governed by the
fluid background. The average energies of the impurity and fluid particles
are asymptotically equal when $m\Theta/T\to 1$, or when $mc=1$.  Thus, energy
equipartition occurs on a particular surface in the three dimensional space
$(m,r_p,r_q)$ where $m(1-q)^2=1-q^2-2p(1-p)$.  Using relations
(\ref{rp})--(\ref{rq}) between the collision parameters and the restitution
coefficients, energy equipartition occurs when the impurity mass $m_{\rm eq}$ is
given by
\begin{equation}
\label{m1}
m_{\rm eq}={1+r_p^2-2r_q^2\over 1-r_q^2}.
\end{equation}
As expected, this mass equals unity when $r_p=r_q$. Curiously, $m_{\rm eq}$ vanishes
when $r_q^2=(1+r_p^2)/2$, indicating that for $r_q>\sqrt{(1+r_p^2)/2}$,
energy equipartition does not occur, as shown in Fig.~2.  However, in a
generic point in the parameter space $(m,r_p,r_q)$ equipartition does break
down \cite{wp,mp}. This is a signature of the dissipative and nonequilibrium
nature of the system.

\begin{figure}
\centerline{\includegraphics[width=8cm]{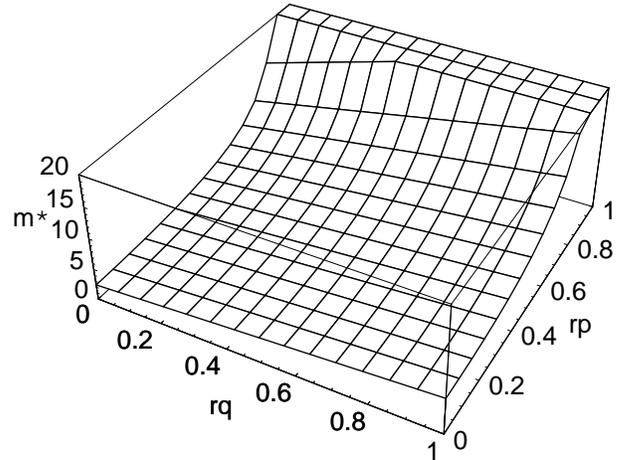}}
\caption{The critical mass $m_*$ versus the restitution coefficients 
$r_p$ and $r_q$.}
\end{figure}

In summary, when the impurity mass is smaller than the critical mass
$m<m_*$, the fluid temperature governs the impurity
temperature. Otherwise, the impurity is infinitely more energetic than
the fluid asymptotically.  As $t\to \infty$ one finds
\begin{equation}
\label{regimes}
{\Theta(t)\over T(t)}\to
\cases{c        &$m<m_*$,\cr
       \infty   &$m\geq m_*$.\cr}
\end{equation}
At the critical impurity mass, $m=m_*$, the solution to
Eq.~(\ref{theta-eq}) shows that the temperature ratio diverges
linearly with time, ${\Theta(t)\over T(t)}\to
2p(1-p)\,t$. Interestingly, the dependence on the dimension is
secondary as it only sets the overall time scale (the transformation
$t\to t/d$ absorbs the dimension dependence).  For example, both the
critical mass, $m_*$, and the temperature ratio $c$ are independent of
$d$.

\begin{figure}
\centerline{\includegraphics[width=8cm]{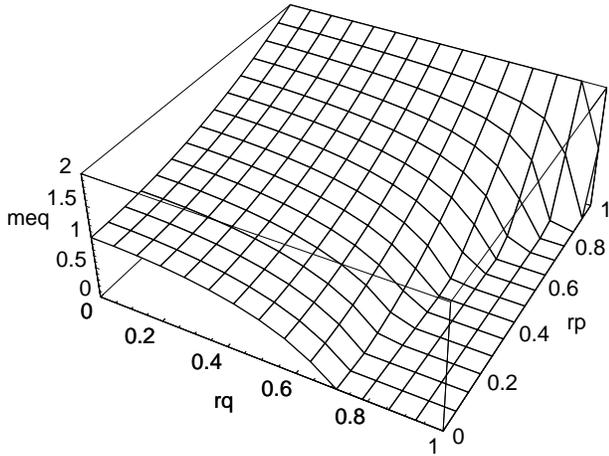}}
\caption{The equipartition mass $m_{\rm eq}$, given by Eq.~(\ref{m1}), 
versus the restitution coefficients $r_p$ and $r_q$.}
\end{figure}

\subsection{The light impurity phase}

For a light impurity, $m<m_*$, the fluid velocity statistics govern basic
characteristics of the impurity statistics. We have seen this for the
temperature, as the initial impurity temperature is irrelevant
asymptotically, and the ratio of the two temperatures approaches a constant
${\Theta(t)\over T(t)}\to c$. The question arises how more detailed
characteristics such as the impurity velocity distribution emerge from the
fluid velocity distribution.  This question can be answered in detail in one
dimension where the explicit scaling solution for the fluid velocity
distribution is known \cite{bmp}.  Below we primarily focus on the
one-dimensional case and derive an explicit exact solution.  We then briefly
comment on the higher-dimensional case, where the solution is very
cumbersome.

The velocity distribution of the fluid approaches a scaling form
asymptotically, $P(v,t)\to T^{-1/2}{\cal
P}\left(vT^{-1/2}\right)$. Following the earlier treatment of the
fluid velocity distribution\cite{bk,bmp,kb,eb}, we employ the Fourier
transform technique.  The fluid Fourier transform is thus
characterized by the scaling form $F(k,t)=f\left(|k|\,T^{1/2}\right)$.
The governing equation\cite{bk} for the corresponding scaling function
is obtained from (\ref{fkt-rate}) to give
\begin{equation}
\label{Fz}
-p(1-p)z f'(z)+f(z)=f(p z)f(z-pz),
\end{equation}
and the solution is $f(z)=(1+z)\,e^{-z}$ \cite{bmp}.

The behavior of the fluid suggests that the impurity velocity distribution
also approaches a scaling form  $Q(v,t)\to T^{-1/2}{\cal
  Q}\left(vT^{-1/2}\right)$. Then the corresponding Fourier transform reads
$G(k,t)=g\left(|k|\,T^{1/2}\right)$. From the Lorentz-Boltzmann equation
(\ref{gkt-rate}), this scaling function satisfies the {\it linear} equation
\begin{equation}
\label{gz}
-p(1-p)z g'(z)+g(z)=(1+az)\,e^{-az}\,g(q z)
\end{equation}
with $a=1-q$. The fluid scaling function is a combination of
$z^ne^{-z}$ with $n=0$ and $n=1$.  We therefore seek a similar
solution of Eq.~(\ref{gz})
\begin{equation}
\label{gsol}
g(z)=\sum_{n=0}^\infty A_n z^ne^{-z}.
\end{equation}
Inserting (\ref{gsol}) into (\ref{gz}), the exponential
factors cancel. Equating terms of the order $z^n$, yields the
following recursion relation for the coefficients:
\begin{equation}
\label{An}
A_n={q^{n-1}(1-q)-p(1-p)\over 1-q^n-np(1-p)}\,A_{n-1}.
\end{equation}
{}From the normalization $g(0)=1$ we get $A_0=1$.  The next two coefficients
are $A_1=1$, $A_2=(1-c)/2$.  Using these values, one computes the small $z$
expansion of the solution (\ref{gsol}): $g(z)=1-cz^2/2+{\cal O}(z^3)$, and
the first non-trivial coefficient is indeed consistent the definition of the
Fourier transform.

The Fourier transform can be inverted to obtain the impurity velocity
distribution function explicitly.  The inverse Fourier transform of
$e^{-\kappa z}$ is ${1\over \pi}{\kappa\over \kappa^2+w^2}$; the
inverse transforms of $z^{n}e^{-z}$ can be obtained using successive
differentiation with respect to $\kappa$. In general, the velocity
scaling function is
\begin{equation}
{\cal Q}(w)={1\over \pi}\sum_{n=0}^{\infty}A_n(-1)^n{d^n\over
d\kappa^n}{\kappa\over\kappa^2+w^2}\Big|_{\kappa=1}.
\end{equation}
In other words, one can express the solution as a
combination of powers of Lorentzians
\begin{equation}
\label{qw}
{\cal Q}(w)={2\over \pi}\sum_{n=2}^{\infty}B_n
\left({1\over 1+w^2}\right)^n.
\end{equation}
The coefficients $B_n$ are linear combinations of the coefficients $A_k$'s
with $k\leq n+1$, e.g., $B_2=1-3A_2+3A_3$ and $B_3=4A_2-24A_3+60A_4$.  The
first squared Lorentzian term dominates the tail of the velocity distribution
${\cal Q}(w)\sim{\cal P}(w)\sim w^{-4}$, as $w\to \infty$. This algebraic
behavior prevails for $w\gg 1$.  Therefore, the impurity has the same
extremal velocity statistics as the fluid.

An interesting aspect of this solution is that for particular
values of the collision parameter $q$, the infinite sum terminates at
a finite order. Of course, when $q_1=p$ then
\begin{equation}
\label{Pw}
{\cal Q}(w)={\cal P}(w)={2\over \pi}{1\over (1+w^2)^2}.
\end{equation}
This is reflected by the vanishing coefficient $A_2=0$. The
multiplicative recursion relation (\ref{An}) then shows that
$A_n=0$ for all $n\geq 2$.  Furthermore, for
$q_2^2(1-q_2)=p(1-p)$, one has $A_3=0$, and the velocity distribution
(\ref{qw}) contains only two terms
\begin{equation}
\label{Pv3}
{\cal Q}(w)={2\over \pi}\left[
{1-3A_2\over\left(1+w^2\right)^2}
+{4A_2\over\left(1+w^2\right)^3}
\right],
\end{equation}
with $A_2=q/(1+2q)$.  Similarly, the solution can be a finite sum with
$k$ terms when $p$ and $q_k$ are related via $q_k^k(1-q_k)=p(1-p)$
thereby imposing $A_n=0$ for $n>k$.  For given fluid collision
parameter $p$, the above relation has the solution $q_k$ only for
sufficiently small $k$.  Thus, there are a few special values of the
impurity collision parameter $q_k$ for which the scaled impurity
velocity distribution is a linear combination of simple rational
functions $\left(1+w^2\right)^{-n}$ with $n=2,\ldots,k+1$.

While the scaling functions underlying the impurity and the fluid are
similar, more subtle features may differ. In particular, the full time
dependent behavior, as characterized by the moments of the impurity
distribution, exhibits rich behavior.  Let $L_n(t)=\int dv\, v^n\,P(v,t)$ be
the moments of the fluid velocity distribution.  Multiplying the equations
(\ref{bep})--(\ref{beq}) by $v^n$ and integrating, the moments obey the
recursive equations
\begin{eqnarray}
\label{l2n}
{d\over dt}L_n+a_nL_n
&=&\sum_{j=2}^{n-2} {n\choose j} p^{j}(1-p)^{n-j}L_{j}L_{n-j},\\
\label{m2n}
{d\over dt}M_n+b_nM_n
&=&\sum_{j=0}^{n-2} {n\choose j} q^{j}(1-q)^{n-j}M_{j}L_{n-j},
\end{eqnarray}
with 
\begin{equation}
\label{abn}
a_n(p)=1-p^n-(1-p)^n, \qquad
b_n(q)=1-q^n.  
\end{equation}
Asymptotically, the fluid moments decay exponentially according to\cite{bk}
\begin{equation}
\label{Ln}
L_{n}(t)\propto e^{-a_n(p)\,t}.
\end{equation}
Using this asymptotics we analyze the behavior of the impurity moments.  The
second moment, i.e. the impurity temperature, was already shown to behave
similar to the fluid temperature when $a_2(p)<b_2(q)$.  The fourth moment
behaves similarly to the fourth moment of the fluid when $a_4(p)<b_4(q)$, and
generally the first $n$ moments are proportional to each other,
\hbox{$M_2\propto L_2,\ldots,M_{2n}\propto L_{2n}$}, when
$a_{2k}(p)<b_{2k}(q)$ for $k=1,\ldots,n$.  In particular, {\em all}
respective impurity and fluid moments are proportional to each other when
$a_{2k}(p)<b_{2k}(q)$ is valid for every $k$.  It is easy to see that above
inequalities hold when $p+q<1$, which is always obeyed when $m<1$.

However, in the complementary parameter range, $a_{n}(p)>b_{n}(q)$ for
sufficiently large $n$.  Such moments are no longer governed by the fluid and
\begin{equation}
\label{mn-heavy}
M_n(t)\propto e^{-(1-q^n)t}.
\end{equation}
The corresponding moment ratio diverges asymptotically: $M_n/L_n\to\infty$ as
$t\to\infty$.  Interestingly, the same behavior (\ref{mn-heavy}) is found in
the heavy impurity phase, as will be shown below. Therefore, the two phases
are not entirely distinct.  A series of transitions affecting moments of
decreasing order occurring at increasing masses,
\begin{equation}
\label{masses}
m_1>m_2>\cdots>m_{\infty},
\end{equation}
signals the transition to the heavy impurity phase.  When $m\geq m_n$, the
ratio $M_{2k}/L_{2k}$ diverges asymptotically for all $k\ge n$. This
generalizes the second moment transition occurring at $m_1\equiv m_*$. The
transition masses 
\begin{equation}
m_n={r_q+{1\over 2}\left[(1-r_p)^{2n}+(1+r_p)^{2n}\right]^{1\over 2n}\over 
1-{1\over 2}\left[(1-r_p)^{2n}+(1+r_p)^{2n}\right]^{1\over 2n}}
\end{equation}
are found from $q^{2n}=p^{2n}+(1-p)^{2n}$ and Eq.~(\ref{rq}).  In
particular, $m_\infty=\lim_{n\to\infty}m_n=(1+r_p+2r_q)/(1-r_p)$, and
sufficiently light impurities ($m<m_\infty$) mimic the fluid
completely.  Additionally, since $m_n>m_\infty\geq 1$, the impurity
must be heavier than the fluid for any transition to occur.

The above analysis of the light impurity phase suggests that in higher
dimensions, the impurity velocity distribution might be similar to
that of the fluid.  In higher dimensions we again assume that the
Lorentz-Boltzmann equation admits a scaling solution.  The
corresponding equation in Fourier space (\ref{gkt-rate}) then
considerably simplifies.  Following the earlier treatments of the
fluid case\cite{kb,eb} we extract the high-energy tail from the
small-$k$ behavior of the Fourier transform of the scaled velocity
distribution. The outcome is that both velocity distributions have the
same algebraic high-energy tail
\begin{equation}
\label{taild}
{\cal Q}(w)\sim{\cal P}(w)\sim w^{-\sigma},
\end{equation}
with the exponent $\sigma$ calculated in \cite{kb,eb}.  Such analysis
also yields the ratio of the prefactors governing this algebraic decay.

\subsection{The heavy impurity phase}

When the mass of the impurity is equal to or larger than the critical
mass, $m\geq m_*$, the velocities of the fluid particles are
asymptotically negligible compared with the velocity of the
impurity. Hence, in the $t\to\infty$ limit, fluid particles become
stationary as viewed by the impurity.  Therefore, one can set ${\bf
u}_2\equiv 0$ in the collision rule (\ref{ruleq}):
\begin{equation}
\label{lorcol}
{\bf v}={\bf u}-(1-q)\,({\bf u}\cdot{\bf n})\, {\bf n}.
\end{equation}
This process is somewhat analogous to a Lorentz gas\cite{Lor}.
However, in the granular impurity system, a heavy particle scatters of
a static background of lighter particles, while in the Lorentz gas the
scatterers are infinitely massive.  Despite this difference, the
mathematical descriptions of the two problems are
similar. Specifically, the collision rule for the inelastic Lorentz
gas \cite{p,mp1} is obtained from (\ref{lorcol}) by a mere replacement
of the factor $(1-q)$ with $(1+r_q)$.

Let us first consider the one-dimensional case where an explicit solution of
the velocity distribution is possible. Setting $u_2\equiv 0$ in the delta
function in the Lorentz-Boltzmann equation (\ref{beq}), integration over the
fluid velocity $u_2$ is trivial, $\int d u_2\,P(u_2,t)=1$, and integration
over the impurity velocity $u_1$ gives
\begin{equation}
\label{lorentz}
{\partial\over \partial t}Q(v,t)+Q(v,t)={1\over q}
Q\left({v\over q}\,,\,t\right).
\end{equation}
This equation can be solved directly by considering the stochastic
process the impurity particle experiences.  In a sequence of
collisions, the impurity velocity changes according to $v_0\to q
v_0\to q^2 v_0\to \cdots$ with $v_0$ the initial velocity.  After $n$
collisions the impurity velocity decreases exponentially, $v_n=q^n
v_0$.  The collision rate is unity, and hence, the average number of
collisions experienced till time $t$ equals $t$. Furthermore, the
collision process is random, and therefore, the probability that the
impurity undergoes exactly $n$ collisions up to time $t$ is Poissonian
$t^ne^{-t}/n!$. Thus, the velocity distribution function reads
\begin{equation}
\label{pivt}
Q(v,t)=e^{-t}\sum_{n=0}^{\infty}{t^n\over n!}\,
{1\over q^n}\,Q_0\left({v\over q^n}\right),
\end{equation}
where $Q_0(v)$ is the initial velocity distribution of the impurity.
Indeed, one can check that this properly normalized solution satisfies
Eq.~(\ref{lorentz}).

Interestingly, the impurity velocity distribution function is a
time-dependent combination of ``replicas'' of the initial velocity
distribution. Since the corresponding argument is stretched, compact velocity
distributions display an infinite set of singularities, a generic feature of
the Maxwell model\cite{e,bk}.

The impurity velocity distribution exhibits interesting asymptotic behaviors.
Consider for simplicity the uniform initial velocity distribution: $Q_0(v)=1$
for $|v|<1/2$ and $Q_0(v)=0$ otherwise.  The solution (\ref{pivt}) reduces to
a finite sum, with $n\leq N={\ln(2v)\over \ln q}$.  In the physically
interesting limits $t\to\infty$ and $v\to 0$, the sum on the right-hand side
of Eq.~(\ref{pivt}) simplifies to respectively $e^{t/q}$ or $(t/q)^N/N!$ when
the number of terms is above or below the threshold value $N=t/q$.  The
magnitude of $Q(v,t)$ above the threshold greatly exceeds the magnitude below
the threshold, so $Q(v,t)$ appears to approach a step function.  A refined
analysis shows that the width of the front widens diffusively so the front
remains smooth although its relative width vanishes.  Specifically, one finds
the following traveling-wave like scaling solution
\begin{equation}
\label{pivt1}
Q(v,t)\to e^{-t+t/q}\,\Psi(\eta),
\end{equation}
with the following wave form and coordinate
\begin{equation}
\label{pivt2}
\Psi(\eta)={1\over \sqrt{\pi}}\int_{-\infty}^\eta dx\,e^{-x^2}, \quad
\eta={{q\over \ln q}\,\ln(2v)-t\over
\sqrt{2qt}}.
\end{equation}
Note, however, that the large velocity tail ($\eta\to\-\infty$),
ignored in Eq.~(\ref{pivt2}), provides actually the dominant
contribution to the moments. This is an unusual traveling wave
form in the sense that the argument is the logarithm of the
velocity rather then the velocity itself, a reflection of the
exponentially decaying velocity.

In contrast to the velocity distribution, the moments $M_n(t)=\int dv\, v^n
Q(v,t)$ exhibit a much simpler behavior.  Indeed, from Eq.~(\ref{lorentz})
one finds that every moment is coupled only to itself, ${d\over dt}
M_n=-(1-q^n)M_n$.  Solving this equation we recover Eq.~(\ref{mn-heavy}); in
the heavy impurity phase, however, it holds for all $n$.  Therefore the
moments exhibit multiscaling asymptotic behavior.  The decay coefficients,
characterizing the $n$-th moment, depend on $n$ in a nonlinear fashion.  This
multiscaling behavior excludes scaling solutions with sharp tails (stretched
exponentials decays and faster). While collisions with the fluid are
sub-dominant, they still lead to corrections to the leading asymptotic
behavior.

We turn now to arbitrary spatial dimensions $d$.  The impurity velocity
changes according to Eq.~(\ref{lorcol}) with the impact direction ${\bf n}$
chosen randomly.  The corresponding Lorentz-Boltzmann equation reads
\begin{eqnarray}
\label{dlorentz}
{\partial\over \partial t}Q({\bf v},t)+Q({\bf v},t)&=&\int d{\bf n}\int
d{\bf u}\,Q({\bf u},t)\nonumber\\
&& \times\delta\Big[{\bf v}-{\bf u}+(1-q)({\bf u}\cdot{\bf
n})\, {\bf n}\Big].
\end{eqnarray}
Moments of the velocity distribution
\begin{equation}
\label{mnt-def}
M_n(t)=\int d{\bf v}\, v^n\, Q({\bf v},t),
\end{equation}
can be obtained directly. We focus on the even moments of the
distribution.  Indeed, they satisfy the following evolution equation
\begin{equation}
\label{mnt-eq}
{d\over dt}M_{2n}=-\big(1-\langle \xi^n\rangle\big)M_{2n}
\end{equation}
where $\mu=\cos^2\theta=(\hat {\bf u}\cdot {\bf n})^2$, $\xi\equiv
\xi(q,\mu)=1-(1-q^2)\mu$, and $\langle \cdot \rangle$ is the shorthand
notation for the angular integration: $\langle \,f
\,\rangle\equiv\int_0^1 {\cal D}\mu\, f(\mu)$. Since $d{\bf n}\propto
\sin^{n-2}\theta\,d\theta$, the (normalized) integration measure
${\cal D}\mu$ is
\begin{equation}
\label{dmu}
B\left({1\over 2},{d-1\over 2}\right){\cal D}\mu
=\mu^{-{1\over 2}}(1-\mu)^{d-3\over 2} d\mu
\end{equation}
where $B(a,b)$ is the beta function. For example, $\langle 1\rangle =1$, and
$\langle \mu\rangle=1/d$.

{}From the evolution equations (\ref{mnt-eq}), the moments are found
to decay exponentially with time
\begin{equation}
\label{mnt-sol}
M_{2n}(t)=M_{2n}(0)\,e^{-\left(1-\langle \xi^n\rangle\right)t}.
\end{equation}
In particular, $1-\langle \xi\rangle=(1-q^2)/d$, and thus, the temperature
decay $\Theta(t)=\Theta(0)\,e^{-(1-q^2)t/d}$ is recovered. There is a little
discrepancy with the exact temperature of Eq.~(\ref{theta-sol}) which depends
on the initial fluid temperature, $T_0$.  This is a remnant of a transient
regime where the two velocity scales are comparable.  Generally, the time
dependence is correct.  Note also that in one dimension $\xi=q^2$ and hence
Eq.~(\ref{mnt-sol}) reduces to our earlier result. In the infinite dimension
limit, $\mu\to 0$, and all moments decay according to $e^{-t}$.  However, in
general, the moments exhibit multiscaling asymptotic behavior, and knowledge
of the typical velocity is insufficient to fully characterize the entire
velocity distribution.  Indeed, writing $M_{2n}\sim M_2^{\alpha_n}$, the
exponents \hbox{$\alpha_n=(1-\langle\xi^n\rangle)/(1-\langle\xi\rangle)$}
have a nontrivial spectrum.

The moments directly give a formal exact solution of the Fourier
transform of the impurity velocity distribution (\ref{gkt-def}).  We
consider isotropic situations where $G\left({\bf k},t\right)\equiv
G(k^2,t)$ with $k\equiv |{\bf k}|$. Expanding the transform in powers
of $k^2$ and substituting the moment result (\ref{mnt-sol}) yields
\begin{equation}
\label{gkt-sol}
G({\bf k},t)=e^{-t}\sum_{n=0}^\infty
{\left(-k^2\right)^n\langle\mu^n\rangle\over (2n)!}\,M_{2n}(0)\,
e^{\langle \xi^n\rangle t}\,.
\end{equation}
While this is an explicit solution, it is not too illuminating. First,
it is in Fourier space, and second, it involves the complicated
angular averages $\langle\xi^n\rangle$.

Nevertheless, it can be shown that the solution remains a time
dependent combination of properly modified replicas of the initial
distribution.  Indeed, either from Eq.~(\ref{gkt-sol}) or directly
from the Fourier transform equation \hbox{${\partial\over\partial
t}G({\bf k},t)+G({\bf k},t)= \langle G({\bf k}\xi,t)\rangle$}, the
solution can be rewritten in the form
\begin{equation}
\label{fkt-alt}
G(k^2,t)=e^{-t}\sum_{n=0}^\infty  {t^n\over n!}\,G_n(k^2).
\end{equation}
Here, $G_0(k^2)$ is the initial Fourier transform, and the ``building
blocks'' $G_n$ are obtained from a recursive procedure of angular
integration $G_{n+1}(k^2)=\langle G_n(k^2\xi)\rangle$.

\section{The Improved Inelastic Maxwell Model}

In the IMM, the rates for fluid-fluid and fluid-impurity collisions
were identical (and therefore, set to unity for convenience).  For
granular fluids, however, the collision rate is proportional to the
relative velocity \cite{cb,my,neg,dufty}.  Therefore, one can improve
the Maxwell model by replacing the actual collision rate with an {\em
average} collision rate proportional to the average relative velocity.
The simplest choice of the average relative velocity is $\sqrt{\langle
({\bf v}_1-{\bf v}_2)^2\rangle }\propto \sqrt{(T_1+T_2)/2}$.  For
fluid-fluid and impurity-fluid collisions we thus obtain $\sqrt{T}$
and $\sqrt{(T+\Theta)/2}$, respectively.  Thus, different collision
rates multiply the collision integrals in the Boltzmann and
Lorentz-Boltzmann equations (\ref{bep})--(\ref{beq}).  In the fluid
case, this overall prefactor merely affects the time dependence of the
temperature.  As will be shown below, in the impurity case, the above
phase transition is suppressed in the IIMM, although secondary
transitions corresponding to higher order moments remain.

Let us again start with the behavior of the temperature. 
The fluid temperature satisfies
\begin{equation}
\label{t-eq-mf}
{d \over dt}T=-\sqrt{T}\left[{2p(1-p)\over d}\,T\right].
\end{equation}
Solving this equation, we recover Haff's cooling law
$T(t)=T_0[1+t/t_0]^{-2}$, with $T_0$ the initial temperature and
$t_0=d/[p(1-p)T_0^{1/2}]$ \cite{pkh}.

The corresponding rate equation for the impurity temperature $\Theta$
is
\begin{equation}
\label{theta-eq-mf}
{d \over dt}\,\Theta=\sqrt{{T+\Theta\over 2}}
\left[-{1-q^2\over d}\,\Theta +{(1-q)^2\over d}\, T\right].
\end{equation}
Since we are primarily interested in the temperature ratio, $S=\Theta/T$, we
study this quantity directly. It evolves according to 
\begin{eqnarray}
\label{ratio-mf}
{1\over \sqrt{T}}\,{d \over dt}\,S &=&\sqrt{{1+S\over2}}
\left[-{1-q^2\over d}\,S+{(1-q)^2\over d}\right]\nonumber\\ 
&+&{2p(1-p)\over d}\,S.
\end{eqnarray}
In the inelastic Maxwell model, the gain and the loss terms were
comparable, both increasing linearly with $S$. Here, in contrast, the
loss term, which grows as $S^{3/2}$, eventually overtakes the gain
term that grows only linearly with $S$.  Therefore, $S\to c$ where $c$
is the root of the cubic equation
\begin{equation}
\label{root-mf} \sqrt{{1+c\over 2}}\left(c-{1-q\over 1+q}\right)
={2p(1-p)\over 1-q^2}\,c.
\end{equation}
Consequently, there is only one phase, the light impurity phase.  We
note that the ratio $c$ is independent of the spatial dimension
$d$. Intuitively, since the impurity collision rate relatively
increases with the impurity temperature, the impurity energy
dissipation rate increases, thereby limiting the (relative) growth of
the impurity temperature.

\begin{figure}
\centerline{\epsfxsize=8cm \epsfbox{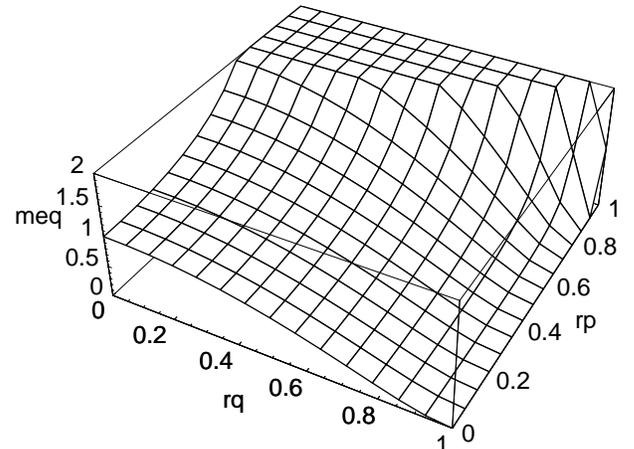}}
\caption{The equipartition mass $m_{\rm eq}$, given by
Eq.~(\ref{m1-mf}), versus the restitution coefficients $r_p$ and
$r_q$.}
\end{figure}

Generally, there is no equipartition of energy except for a particular
surface in the space $(m,r_p,r_q)$. Energy equipartition occurs when
$mc=1$.  Using the relations (\ref{rp})--(\ref{rq}), 
we find the equipartition
mass 
\begin{equation}
\label{m1-mf}
m_{\rm eq}=-{1\over 2}+{1\over 2}
\sqrt{1+8\left({1-r_q^2\over 1-r_p^2}\right)^2}.
\end{equation}
Figure 3 plots $m_{\rm eq}=m_{\rm eq}(r_p,r_q)$.

Qualitatively, our findings in the light impurity phase of the IMM extend to
the IIMM. For example, both velocity distributions follow scaling forms and
the large-velocity tails of both distributions are the same.  In the one
dimensional case, explicit expressions for the impurity scaling function are
possible, and as the treatment follows closely that outlined in the light
impurity phase, we briefly outline the results.  In 1D, the impurity velocity
distribution approaches a scaling solution $Q(v,t)\to T^{-1/2}{\cal Q}\left(v
  T^{1/2}\right)$.  The corresponding Fourier transform reads
$G(k,\tau)=g\left(|k|\,T^{1/2}\right)$. The only difference with the above
inelastic Maxwell model is that the collision terms are proportional to
$\beta^{-1}=\sqrt{(1+c)/2}$.  Consequently, Eq.~(\ref{gz}) generalizes as
follows
\begin{equation}
\label{gz-mf}
-\beta p(1-p)z g'(z)+g(z)=(1+az)\,e^{-az}\,g(q z).
\end{equation}
Seeking a series solution of the form (\ref{gsol}), leads to the following
recursion relations for the coefficients
\begin{equation}
\label{An-mf}
A_n={(1-q)q^{n-1}-\beta p(1-p)\over 1-q^n-n\beta p(1-p)}\,A_{n-1},
\end{equation}
with $A_0=A_1=1$.  Again, the velocity distribution is a combination
of powers of Lorentzians as in Eq.~(\ref{qw}):
\begin{equation}
\label{qw-mf}
{\cal Q}(w)={2\over \pi}\sum_{n=2}^{\infty}B_n
\left(1+w^2\right)^{-n},
\end{equation}
where $B_n$ are linear combinations of the coefficients $A_n$'s given
by the same expressions as in Sec.~II. In particular, the
large-velocity tail is generic ${\cal Q}(w)\sim w^{-4}$.

Given the algebraic form of the velocity distributions, we examine the
asymptotic behavior of moments of the velocity distribution.  Moments
of the fluid and the impurity, $L_n$ and $M_n$, respectively, evolve
according to a straightforward generalization of
Eqs.(\ref{l2n})--(\ref{m2n}),
\begin{eqnarray}
\label{ln-mf}
{d\over d\tau_p}L_n+a_nL_n
&=&\sum_{j=2}^{n-2} {n\choose j} p^{j}(1-p)^{n-j}L_{j}L_{n-j},\\
\label{mn-mf}
{d\over d\tau_q}M_n+b_nM_n
&=&\sum_{j=0}^{n-2} {n\choose j} q^{j}(1-q)^{n-j}M_{j}L_{n-j}.
\end{eqnarray}
Here $a_n(p)$ and $b_n(q)$ are given by Eq.~(\ref{abn}) and  
the collision counters, 
\begin{eqnarray*}
\tau_p=\int_0^t dt'\,\sqrt{T}, \qquad
\tau_q=\int_0^t dt'\,\sqrt{(T+\Theta)/2}, 
\end{eqnarray*}
play the role of time in Eq.~(\ref{ln-mf}) and (\ref{mn-mf}),
respectively.  Since the two temperatures are asymptotically
proportional to each other, $\tau_q\to\tau_p/\beta$.  The fluid
moments decay according to $L_n\propto e^{-a_n(p)\tau_p}\propto
t^{-2a_n/a_2}$ \cite{bk}.  Inserting the asymptotics $L_n\propto
e^{-a_n(p)\beta\tau_q}$ into Eq.~(\ref{mn-mf}) and performing the same
analysis as in the IMM we find that when $\beta a_n(p)<b_n(q)$, the
impurity moments are enslaved to the fluid moments, i.e., $M_n\propto
L_n$ asymptotically. Otherwise, sufficiently large impurity moments
behave differently than the fluid moments, viz. $L_n\propto
e^{-b_n\tau_q}\propto t^{-2b_n/\beta a_n}$.  Although the primary
transition affecting the second moment does not occur, secondary
transitions affecting larger moments do occur at a series of masses,
as in Eq.~(\ref{masses}). In the IIMM $m_1\equiv m_*$ diverges, but
other masses remain finite.  The transition masses $m_n$ are found by
solving $\beta a_{2n}(p)=b_{2n}(q)$ simultaneously with
Eq.~(\ref{root-mf}), and then applying Eq.~(\ref{rq}). For example,
for completely inelastic collisions ($r_p=r_q=0$) one finds
$m_2=1.65$.  Qualitatively, these transitions imply that some velocity
statistics of the impurity, specifically large moments, are no longer
governed by the fluid.

\section{Summary}

We have studied dynamics of impurities in granular fluids using the
inelastic Maxwell model.  In general, there is breakdown of energy
equipartition as two different temperatures characterize the impurity
and the fluid. We analyzed two different models.  First, we considered
identical fluid-fluid and fluid-impurity overall collision rates. In
this case a phase transition, marked by a dissipation dependent
critical impurity mass, occurs.  Breakdown of equipartition is
moderate in one phase and extreme in the other with the asymptotic
temperature ratio diverging. Sufficiently light impurities are
governed by the fluid, and their scaled velocity distribution has
similar extremal statistics as has the fluid.  In one dimension, the
scaled impurity velocity distribution is given by a series containing
powers of Lorentzians of all orders (the infinite sum truncates for a
few special values of the impurity restitution coefficient).
Interestingly there is a series of increasing transition masses that
corresponds to divergence of the ratio of velocity moments of
decreasing order. These masses are always larger than unity and the
largest such mass corresponds to the temperature.  When the impurity
mass is larger than this critical mass, the impurity and the fluid
effectively decouple. Next, we considered collision rates that account
for the temperature difference between the fluid and the impurity. In
this case, the primary transition corresponding to the temperature is
suppressed but further transitions corresponding to divergence of
higher moments may still occur.

Many more quantities should be tractable in the framework of the
Maxwell model. For example, one can study velocity correlations as
well as the velocity autocorrelation functions. Furthermore,
corrections to the leading asymptotic behavior, especially in the
heavy impurity phase, can be systematically evaluated. Additionally,
this model can be applied to particles of different sizes by modifying
the collision cross sections. 

We reiterate that the inelastic Maxwell model exactly describes only a
mean-field collision process where the collision partners are random
and the impact directions are chosen according to a uniform
distribution. On the other hand, it is an uncontrolled approximation
when applied to real granular fluids. At best, this model may be used
to approximate sufficiently low velocity moments. Still, these results
show that the relative collision rates play an important
role. Comparison with the hard sphere Boltzmann equation, Molecular
Dynamics simulations, and experiment is needed to gauge the utility of
either of the two approximate models we treated.  In practice, the
impurity problem involves an additional level of difficulty. Gathering
meaningful statistics requires many replica systems with one impurity
or alternatively, the vanishing volume fraction limit of a mixture.

\medskip\noindent
This research was supported by
DOE (W-7405-ENG-36), NSF(DMR9978902), and ARO (DAAD19-99-1-0173).

\end{multicols}
\end{document}